# Implementation Science–Driven Identification of Barriers to AI Diagnostic Decision Support: A Stakeholder-Informed Concept Mapping Study

Ali Fathi Jouzdani, Arman Gorji, Mohammad Zahedi, Morteza Alizadeh, Babak Saboury, Arman Rahmim, *Senior Member, IEEE,* Mohammad Salmanpour, and Ahmad Fayaz-Bakhsh

[1]

***Abstract*****—** The clinical translation of artificial intelligence (AI) is a critical step toward precision medicine. The implementation study presented here focuses specifically on barriers to AI adoption in the diagnostic medicine workflow. Successful clinical adoption of AI requires more than technical performance; it depends on addressing real-world organizational, ethical, infrastructural, and workflow-related barriers that may prevent promising systems from moving beyond research settings. To examine these barriers, we used structured concept mapping (CM) to operationalize stakeholder engagement and to incorporate an integrated knowledge translation (iKT) approach within the Knowledge-to-Action (KTA) framework. Stakeholders co-generated 96 statements describing barriers to AI implementation and organized them into 11 clusters, including infrastructure limitations, ethical and legal concerns, workflow integration challenges, and distrust in algorithmic reasoning. These findings provide actionable knowledge for supporting AI implementation in the diagnostic chain. The identified barriers may also have implications for emerging applications, such as digital twins, that rely on AI-driven clinical data and predictive models. Addressing these barriers may help support the integration of AI-enabled diagnostic tools into routine clinical practice and contribute to future advances in data-driven healthcare.

## I. INTRODUCTION

The field of implementation science (IS) provides significant tools and strategies to enable creators and users of knowledge to bridge the know-do gap (innovation-implementation chasm) [1]. With the advent of new digital technologies, including artificial intelligence (AI) in medicine, implementation barriers can pose significant challenges to timely and full uptake of innovations and their transformation into scalable, context-sensitive solutions [2, 3]. As the pace of health innovation accelerates, IS-based strategies and solutions, such as integrated knowledge translation (iKT), provide structured solutions to address the multifaceted challenges of adoption and integration, to improve indicators of implementation outcomes [4]. Importantly, these approaches not only enhance the sustainability of innovations but also ensure that evidence-based practices are equitably translated across diverse healthcare settings [5]. By embedding IS into the innovation cycle, researchers and practitioners can align discovery with real-world impact more effectively [6].

AI has growing potential to support diagnostic medicine by improving risk assessment, workflow efficiency, image and test interpretation, report generation, and communication of results. The diagnostic chain workflow includes test ordering, scheduling, protocoling, acquisition, interpretation, report generation, and report communication [7]. These steps are central to diagnostic specialties such as radiology, nuclear medicine, pathology, hematology, and laboratory medicine. Because diagnostic decisions often shape subsequent clinical pathways, the effective implementation of AI in this workflow may influence both care quality and system performance [7-17].

AI-enabled diagnostic workflows may also contribute to the development of emerging healthcare applications that use patient-specific data and computational models to support prediction and personalized decision-making [18-21]. Accordingly, this study focuses on barriers to AI adoption across the diagnostic chain workflow, while recognizing that effective implementation of AI may have broader implications for future data-driven healthcare applications. Figure 1

[1]This work was supported in part by the Natural Sciences and Engineering Research Council of Canada (NSERC) Discovery Horizons Grant DH-2025-00119. (Corresponding authors: Mohammad Salmanpour, e-mail: msalman@bccrc.ca; Ahmad Fayaz-Bakhsh, e-mail: fayaz@tums.ac.ir; Mohammad Salmanpour and Ahmad Fayaz-Bakhsh also contributed equally as co-last authors.)

This work did not involve human subjects or animals in its research.

Ali Fathi Jouzdani, Arman Gorji, and Mohammad Zahedi are with the NAIRG, Department of Neuroscience, Hamadan University of Medical Sciences, Hamadan 6517838636, Iran (e-mail: ali.fathi77@gmail.com; gorjiarman@gmail.com; zahedimohammad676@gmail.com).

Morteza Alizadeh is with the Department of Mathematics, University of Isfahan, Isfahan 817467344, Iran (e-mail: alizadehmorteza2020@gmail.com).

Babak Saboury is with the Institute of Nuclear Medicine, Bethesda, MD 20814 USA (e-mail: bsaboury@inm.org).

Arman Rahmim is with the Department of Basic and Translation Research, BC Cancer Research Institute, Vancouver, BC V5Z 1L3, Canada, and also with the Department of Radiology, University of British Columbia, Vancouver, BC V6T 1Z4, Canada (e-mail: arman.rahmim@ubc.ca).

Mohammad Salmanpour is with the Department of Basic and Translation Research, BC Cancer Research Institute, Vancouver, BC V5Z 1L3, Canada. He is also with the Department of Radiology, University of British Columbia, Vancouver, BC V6T 1Z4, Canada, and with the Technological Virtual Collaboration (TECVICO CORP.), Vancouver, BC V6E 2E9, Canada (e-mail: msalman@bccrc.ca).

Ahmad Fayaz-Bakhsh is with the Department of Disaster and Emergency Health, School of Public Health, Tehran University of Medical Sciences, Tehran 1417613151, Iran, and also with the Centre for Health Services and Policy Research, University of British Columbia, Vancouver, BC V6T 1Z3, Canada (e-mail: fayaz@tums.ac.ir).

Supplementary material for this article, including the full questionnaire, demographic survey, generated statements (Table S1), and Go-Zone plots (Fig. S1), is available at: https://github.com/MohammadRSalmanpour/Knowledge-to-Action-Approach. The repository also provides access to the analysis codes and data used in this study.

illustrates the AI-enabled diagnostic chain workflow, which is the primary focus of the present study.

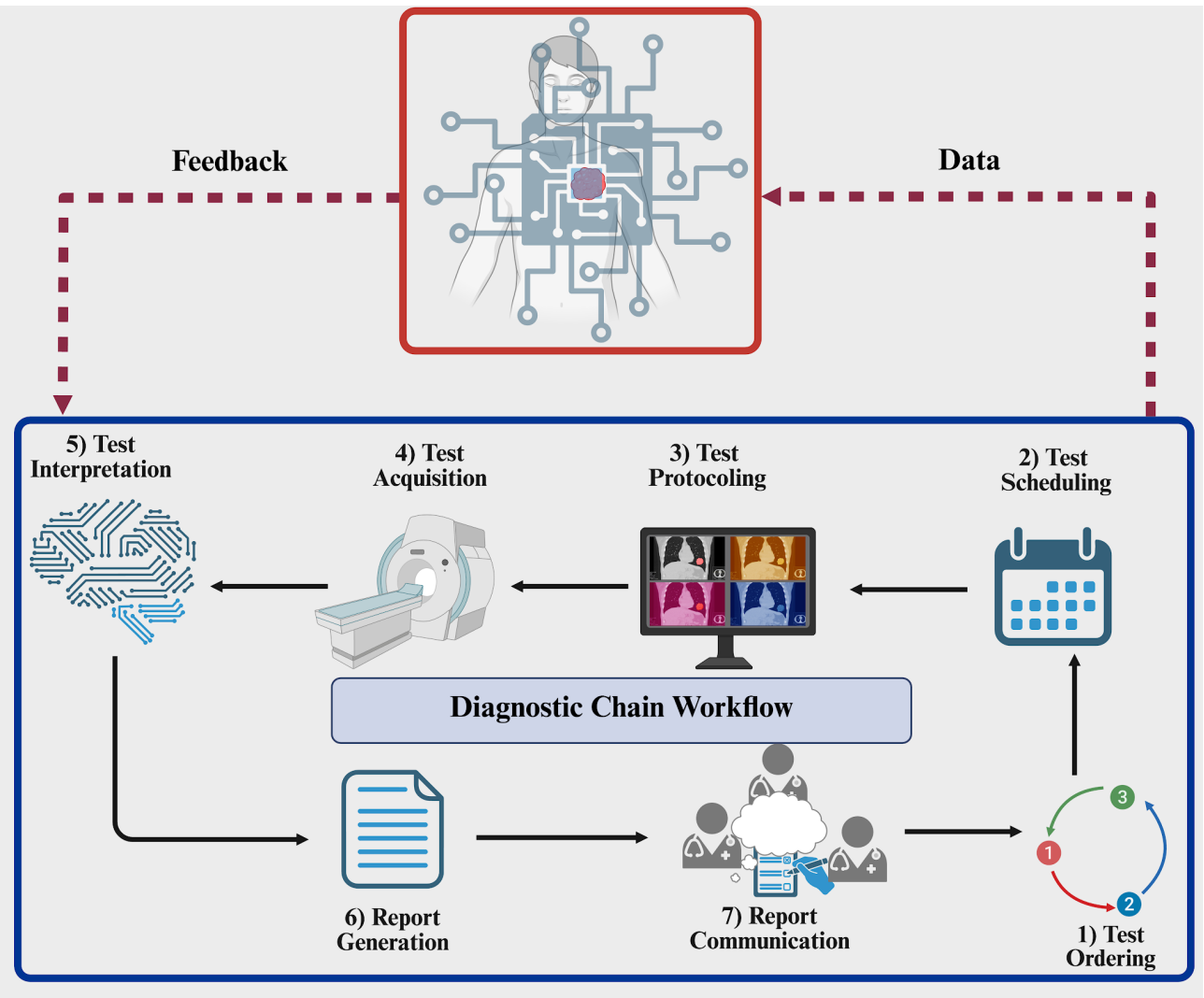

**Fig. 1.** AI-enabled diagnostic chain workflow. The diagnostic chain comprises seven steps: test ordering, scheduling, protocoling, image/test acquisition, interpretation, report generation, and report communication. AI may support these steps by improving diagnostic decision-making, enhancing workflow efficiency, and facilitating the generation, interpretation, and communication of clinical data. The figure illustrates the potential integration of AI across the diagnostic chain workflow to support clinical practice and improve healthcare delivery.

Despite increasing interest in AI for diagnostic medicine, routine implementation remains limited [22]. Reported challenges include data quality concerns, technical integration, limited interpretability, clinician trust, ethical and legal uncertainty, governance issues, and potential disruption to professional roles [23-28]. Implementation may also be affected by variation in digital literacy, communication practices, adaptability, and stakeholder engagement among clinicians, developers, leaders, patients, and caregivers [19, 29-32].

These challenges occur across individual, organizational, institutional, system, and societal levels, and they may affect healthcare settings regardless of resource level [33-35]. Previous implementation studies suggest that stakeholder participation can improve the relevance, feasibility, and acceptability of complex health technologies [35-39]. For AI in diagnostic workflows, early engagement of knowledge users and knowledge creators may be particularly important to ensure that tools are aligned with clinical needs, local workflows, governance requirements, and patient priorities.

Building on a previous analysis [1], we highlight in this work how IS can illuminate translational pathways, foster stakeholder engagement, and address systemic barriers that hinder the adoption of complex interventions. We extend these insights by leveraging a stakeholder-informed concept mapping (CM) exercise aligned with the Knowledge-to-Action (KTA) framework. This approach not only reveals actionable strategies for facilitating implementation but also highlights how IS can bridge innovation and sustainable clinical impact. Specifically, we pursued a participatory mixed-methods approach that captures stakeholder knowledge, supports collaborative interpretation, and facilitates the co-creation of actionable solutions [40]. Our study focused on barriers to AI adoption across the diagnostic chain workflow in clinical medicine, structured according to the seven-step chain workflow model described by Kapoor et al. [7] and illustrated in Fig. 1. Through stakeholder engagement in idea co-generation, sorting, rating, and interpretation, we aimed to identify implementation obstacles and propose evidence-informed, context-sensitive strategies to support AI co-creation and collaborative use among knowledge creators and knowledge users. By applying CM as an IS and iKT approach, we sought to identify stakeholder-perceived barriers to implementing AI across the diagnostic chain workflow towards informed stakeholder-driven implementation strategies. The findings may help inform stakeholder-driven strategies for implementing AI across the diagnostic chain workflow.

## II. METHODS AND MATERIALS

### *A. Framework Overview*

Our project followed the KTA framework [41], which contains two main components: knowledge and action. The framework has been widely adopted in integrated iKT and IS activities, including applications in cancer screening uptake [41]. A systematic review of KTA applications demonstrates its flexibility for linking knowledge co-creation with real-world co-implementation in several contexts [42], including in many medical and public health programs. In the KTA model, knowledge co-creation progresses through a funnel from primary research and synthesis to development of a set of locally applicable evidence-based interventions (EBI). The action cycle includes identifying priority problems, collaborative selection of the most important and feasible from among the locally developed EBIs, co-planning for their uptake, co-assessing the barriers to their implementation, co-applying implementation strategies, and finally collective monitoring and evaluation of implementation strategies and targeted outcomes, and sustaining them [41-43]. In our study, a transformation in participants' mindsets—towards iKT principles and prioritizing co-implementation—was both an intentional procedural step and an outcome measure. Therefore, some procedural details described below inherently reflect early-stage impacts of our implementation process.

An implementation expert guided our team to adopt this iterative model and, through his knowledge brokering strategies, enabled participating stakeholders to progress in their co-production of knowledge on key implementation barriers. Targeted stakeholders in our exercise included clinicians and AI developers who were sufficiently aware of factors contributing to AI underutilization in clinical settings at BC Cancer. Early stakeholder engagement was facilitated by our knowledge translation (KT) expert, whose primary role was not only to build consensus on a collective approach to data collection and analysis, as well as the co-creation of local innovations and EBIs, but also to guide the study participants in their co-implementation of locally developed knowledge. Our KTA and CM helped us co-identify barriers, and we plan to use this insight to inform the subsequent development of EBIs that are most important and feasible to implement locally. In line with this, we will undertake steps for knowledge co-creation on AI, co-identifying priorities among AI tools/systems developed in our lab, co-design, and co-implementation of our plans in the future. What is reported here is about a part of our co-implementation activities, which were integrated into the post-planning phase. This paper also provides a detailed report on the entire steps taken by the knowledge broker, creators, and users during the pre-implementation phase in this exercise. More detailed reporting on full implementation and sustainment phases will be presented in forthcoming publications.

### *B. Methods–Results Integration in IS*

In IS, it is often impossible to fully separate methodological descriptions from study results, because many collaborative activities generate both procedural insights and tangible outcomes. This overlap is not unique to our study; others have reported similar integration of methods and results [44, 45]. For example, activities such as co-creation, co-dissemination, co-uptake, co-implementation, co-sustainment, and co-evaluation are simultaneously methodological processes and signs of success measurable by indicators for various implementation outcomes. In our study, a transformation in participants' mindsets—towards iKT principles and prioritizing co-implementation—was both an intentional procedural step and a measure of implementation outcome. Therefore, some procedural details described below inherently reflect early-stage impacts of our implementation process.

### *C. Seven Steps of the CM Approach*

As stated earlier, we used CM, which is a mixed-methods approach and allows for integration of qualitative and quantitative methods in data collection, in data analysis, and in implementation (e.g., multidimensional scaling (MDS), hierarchical cluster analysis) [40, 46]. CM structures stakeholder-generated ideas into visual maps of conceptual domains and is well-suited to KTA-based IS research. As shown in Fig. 2, we followed the standard seven steps of the CM. We adapted these steps to the context of AI implementation in clinical diagnostic workflows, focusing on barriers to AI adoption across the seven stages of the diagnostic chain illustrated in Fig. 1.

Step 1: Preparation. We recruited 36 stakeholders, including end-users (e.g., physicians and frontline clinicians), clinical AI researchers, implementation experts, digital health strategists, and health system leaders. Participants were purposefully selected to ensure diversity in expertise, sector representation, and demographics, thereby enabling a comprehensive perspective on the integration of AI into clinical diagnostic workflows. Recruitment was conducted using snowball sampling and community-based outreach, targeting individuals with experience in healthcare delivery, clinical workflow operations, or AI system development. During this step, preparation was made for the next steps in terms of how to perform further participant recruitment until saturation is reached, brainstorming and statement generation, statement selection and revisions, sorting and rating, and so on.

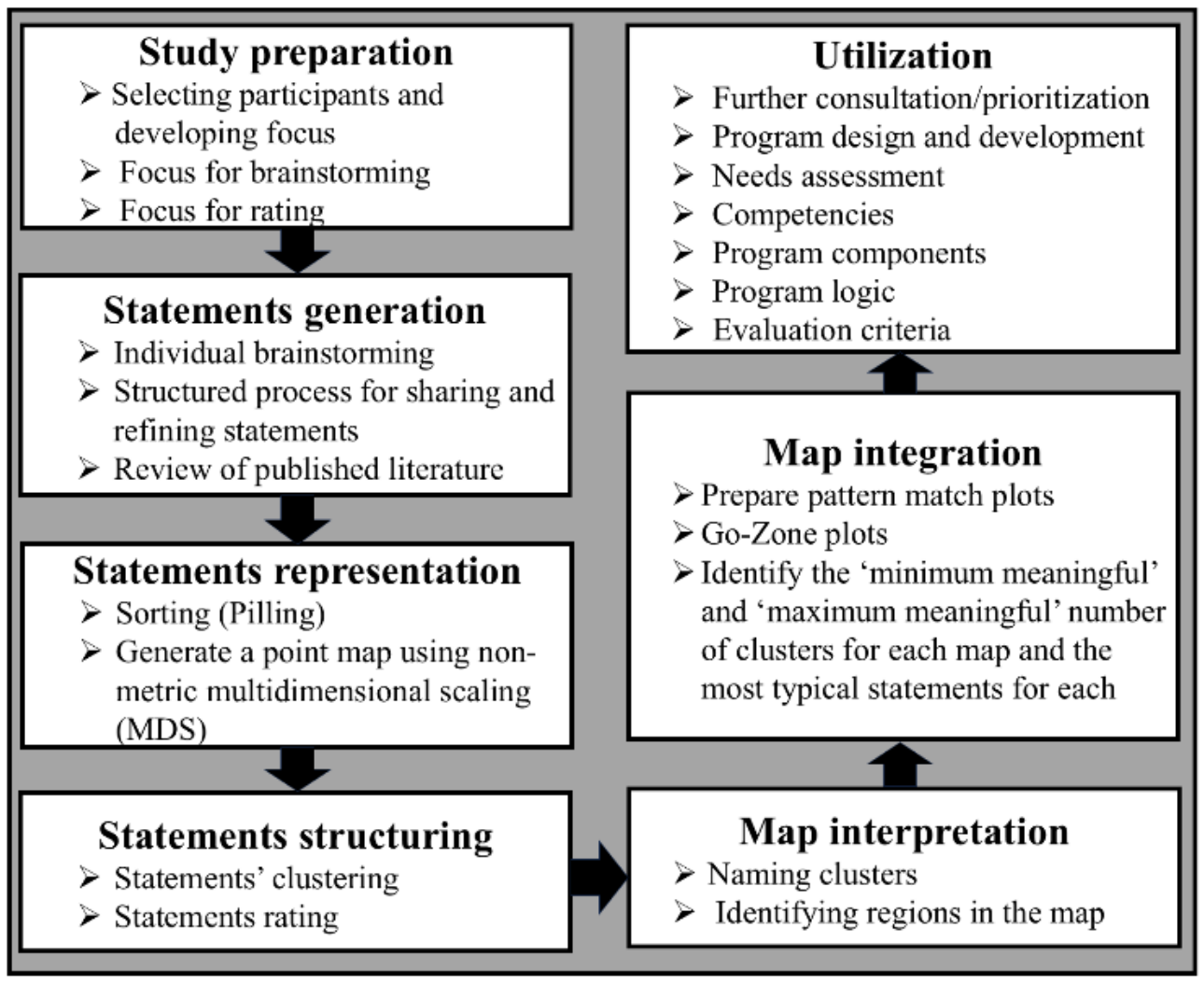

**Fig. 2.** The Conceptual Mapping structured process for generating, structuring, and interpreting statements through a series of steps, culminating in the utilization of maps for prioritization and program design.

Step 2: Statement Generation. Following the CM methodology, statement generation was initiated through engagement with recruited stakeholders. A collaboratively developed focus prompt— "What barriers or challenges exist for integrating AI tools in the diagnostic chain workflow?"—was used to guide suggestions of several statements as our initial input in this step. To ensure relevance and context specificity, the prompt was embedded in a scenario-based questionnaire representing seven steps of the diagnostic process. Stakeholders provided open-ended responses, which were compiled into an initial statement pool. Two researchers independently reviewed the statements, removing redundancies, clarifying wording, and ensuring each statement expressed a single coherent idea. Relevant literature was then consulted to add statements, to verify completeness, and to maintain alignment with established scientific domains and paradigms. This was then succeeded by team efforts to finalize the generation and agree on a set that included 96 statements. This set, as we explain below, was used for our next CM activities.

Step 3: Statement Representation. In accordance with CM methodological guidelines, representation activities centered on the analysis of sorting data. We first constructed individual participant similarity matrices, which were then aggregated into a total similarity matrix (TSM). Using the TSM, point maps were generated through non-metric MDS implemented in Python (v3.10), as mentioned in the Code and Data Availability section. These MDS-based point maps served as a key tool for visualizing conceptual diversity and proximity between statements. To identify clusters, participants—engaging either virtually or in person—examined the spatial arrangement of points and discussed the most appropriate way to draw the minimal number of polygon-shaped boundaries, ensuring each point was enclosed within exclusively one cluster and no point was left out of the boundaries. In line with IS and iKT principles, an implementation expert facilitated these discussions, promoting consensus on the optimal cluster structure. The hierarchical clustering process was iterative, with refinement continuing until thematic groupings achieved collective agreement. The results of this section were subsequently used to produce the final cluster map.

Step 4: Statement Structuring. In accordance with the CM approach, the finalized statement set was structured through two participant-led activities: sorting and rating. As mentioned above, in our sorting task, participants grouped statements into categories that reflected their own perceptions of thematic similarity, creating clusters without a predetermined framework. This process allowed for the emergence of participant-defined conceptual groupings. To achieve the structuring of statements based on some rating data, we asked our participants to independently assess each statement along predefined dimensions—such as importance and feasibility—using a Likert-type scale. Sorting data was later analyzed to generate similarity matrices for MDS and hierarchical cluster analysis, while rating data were retained for subsequent interpretation of cluster-level priorities.

Step 5: Map Interpretation. Based on methodological guidelines, the interpretation activities in any CM exercise should primarily focus on "labeling"—facilitating participants in reaching widely agreed names for the polygon-shaped boundaries, known as cluster on CM point maps. This step is essential for knowledge co-creation, integrated iKT, and the co-implementation of knowledge within the CM framework. To achieve this, participants from diverse stakeholder groups were supported to review the cluster map, interact, and collectively agree on the most appropriate titles for each cluster. Through iterative and continuous interaction, our knowledge broker mediated and guided the process and helped participants to refine the proposed labels for different clusters.

Step 6: Map Integration. In the final step of the CM process, the cluster map and associated rating data were integrated to create a comprehensive visual representation of the way forward in AI integration activities in our laboratories. For example, to come up with a list of the most feasible and important set of integration strategies, we calculated the mean rating values for each cluster. We also plotted a series of pattern match and go-zone figures as recommended in the CM methodology. The plot of pattern match graphs enabled comparisons at the cluster-level based on average ratings between AI users and AI developers across importance and other dimensions, highlighting areas of alignment and divergence among different rating criteria. Go-Zone plots helped us to identify statements rated highly according to various pairings of importance with other rating dimensions. Together, these tools provided complementary insights: Pattern Match offered a macro-level view of stakeholder alignment, while Go-Zone plots delivered micro-level, actionable priorities. This integrated map and its visual outputs formed the foundation for synthesizing stakeholder perspectives into an actionable framework to address barriers to AI integration in diagnostic workflows during the utilization step.

Step 7: Utilization. Based on the final methodological guidelines of CM, the last set of activities focuses on utilization, which encompasses multiple steps within the action component of the KTA framework. This step is dedicated to disseminating and promoting the uptake of knowledge generated locally. In our exercise, for example, we focused on KT regarding barriers to the adoption of evidence-based AI in future efforts. Details of this action, along with the implementation component of our KTA and IS exercise, will be reported separately in the next report. Moving forward, we plan to use these findings to inform the design and deployment of tailored, context-sensitive implementation strategies for integrating AI into clinical diagnostic workflows, particularly in our affiliated AI laboratories and diagnostic clinics. During the future utilization step, the primary objective will be to leverage the results of preceding activities to guide the selection, adaptation, and implementation of interventions (Fig. 2).

## III. RESULTS

### *A) Study Preparation Results*

A total of 36 stakeholders participated in the study, representing three broad groups: end-users, including physicians and frontline clinicians; clinical AI researchers and implementation experts; and digital health strategists and health system leaders. Of these, 14 participants contributed to the semi-structured interviews used to generate the initial statement set. Subsequently, 12 participants completed the statement sorting and point-clustering tasks, and 10 participants completed the statement rating exercises.

The rating sample included five AI users and five AI developers, all of whom had completed the preceding study tasks. The AI user group comprised physicians (n = 3) and radiology technicians (n = 2), all of whom had direct experience using AI-enabled tools in healthcare practice. The AI developer group comprised data scientists (n = 2), a mathematician (n = 1), and technical experts involved in the development, testing, or evaluation of AI-based systems (n = 2). Participants reported between 7 and 15 years of professional experience in their respective domains. The rating exercises were conducted across five predefined dimensions: (i) "Reason to Use" – barriers related to core legal, financial, or similar obligations for adopting AI tools in clinical care, (ii) "Means to Use" – barriers related to required skills, tools, or infrastructure for AI adoption, (iii) "Method to Use" – barriers related to practical or acceptable methods for engaging with AI in routine workflows, (iv) "Desire to Use" – barriers related to individual, organizational, or other motivational factors supporting AI adoption [47], and (v) "Importance" – the significance of the barrier in preventing successful and sustainable AI implementation [48]. These dimensions were embedded in a 26-item scenario-based questionnaire relating to seven steps of the diagnostic workflow (e.g., test ordering, test interpretation, test protocoling) [7, 11]. These scenarios facilitated context-specific reflections on barriers (Supplementary Material, Section 1). Responses collected via Google Forms yielded an initial pool of statements. Two researchers then applied collaborative coding, refinement, and reduction processes to remove redundancies and clarify wording.

### *B) Statement Generation Results*

Stakeholder engagement produced a diverse range of perspectives on barriers to AI integration, reflecting both technical and human-centered challenges. The iterative review process yielded 96 distinct, well-defined statements spanning all seven steps of the diagnostic process. These statements addressed issues such as workflow integration gaps, trust and accountability concerns, adaptability limitations, and ethical or legal challenges. The complete list of finalized statements is provided in Supplemental File, Section 2 (Supplemental Table S1), and served as the foundation for subsequent sorting and rating activities in the CM process.

## *C) Statement Representation*

The two-dimensional point map generated through non-metric MDS visually depicted the relational structure among the 96 statements (Fig. 3). Each point represented a single statement, and the distance between points indicated the frequency with which participants grouped them during the sorting task. Statements located closer together were more frequently placed in the same pile, reflecting higher perceived thematic similarity, whereas those farther apart were less frequently associated. This spatial configuration formed the basis for hierarchical cluster analysis, which identified coherent AI barrier clusters relevant to integration within diagnostic workflows.

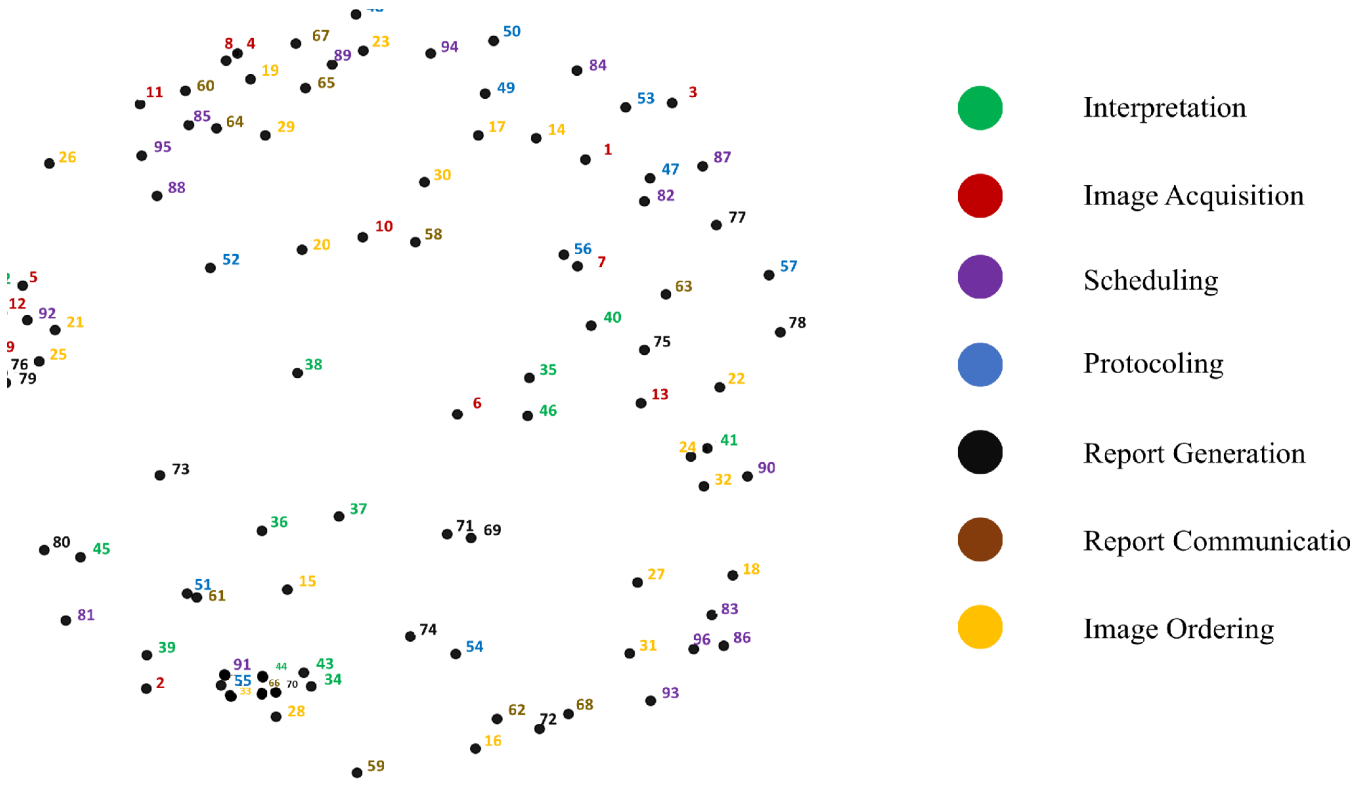

**Fig. 3.** Two-dimensional point map of 96 statements generated from participant sorting data. Each point represents one of the 96 statements. The distance between the points on the above map was affected by how often our participants agreed on their similarities. In the case of points with closer proximity, it reflects greater frequency in perception by our sorting participants of their thematic similarity.

## *D) Statement Structuring and Map Interpretation Results*

After the representation steps, a point map was generated from the statement representations and served as the basis for clustering. Participants' collective and close scrutiny of points depicted on Fig. 3 resulted in the identification of 11 distinct clusters representing key perceived barriers to AI in diagnostic workflows. These clusters, shown in Fig. 4, capture the multidimensional landscape of stakeholder concerns and offer a structured foundation for implementation planning.

(i) Infrastructure and Integration Barriers: Challenges related to inadequate hardware, software, and interoperability infrastructure are required to support AI systems in clinical environments. (ii) Clinical Skepticism and Loss of Human-Centered Decision-Making: Concerns that AI tools may undermine the clinician's role or erode the humanistic aspects of medical judgment and patient care. (iii) Insufficient Adaptability to Clinical Complexity: Perceptions that AI systems are not sufficiently flexible to accommodate the diverse needs, workflows, and patient presentations encountered in real-world clinical practice. (iv) Trust and Accountability Issues in AI-Generated Reporting: Apprehensions about the reliability, traceability, and transparency of reports generated by AI systems, particularly in high-stakes diagnostic contexts. (v) Ethical, Legal, Financial, and Privacy Concerns: Worries related to patient data security, legal liabilities, ethical implications, and the cost-benefit justification of implementing AI technologies. (vi) Lack of Training and Familiarity with AI Tools: Deficiencies in formal training and limited user familiarity contribute to discomfort, misinterpretation, and resistance among clinicians and staff. (vii) Workflow Disruption and Lack of Standardization: Disruption to established diagnostic processes and the absence of standardized procedures for incorporating AI into routine workflows. (viii) Workforce Resistance Due to Job Security Concerns: Fears that AI could lead to workforce displacement, reduced autonomy, or professional devaluation among clinical staff. (ix) Limitations of AI in Handling Complex or Atypical Cases: Skepticism about AI's ability to navigate rare, ambiguous, or high-variability clinical situations where human expertise is often essential. (x) Clinical Distrust and Concerns over AI Decision-Making: Lack of confidence in the clinical reasoning and decision-making capabilities of AI systems, particularly in nuanced or uncertain diagnostic situations. (xi) Bias, Reliability, and Quality Challenges in AI Performance: Concerns about algorithmic bias, inconsistent outputs, and variability in AI performance that may reinforce health disparities or compromise patient safety.

Together, these 11 clusters represent a comprehensive map of the perceived obstacles to AI diagnostic decision support adoption. They reflect systemic, technical, behavioral, and ethical dimensions of implementation and underscore the need for targeted, stakeholder-informed strategies to address these barriers in a context-sensitive manner.

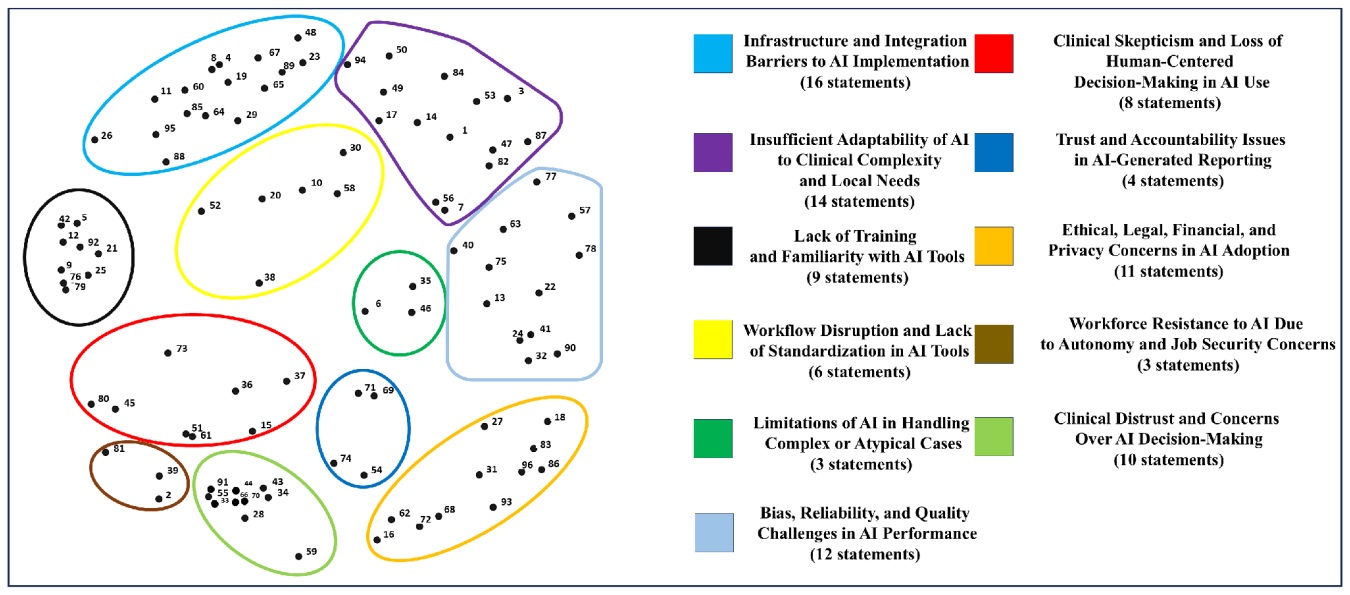

**Fig. 4.** Concept map representing the barriers to AI adoption in clinical settings, with clusters highlighting key challenges. Each color-coded group corresponds to a specific area of concern.

The thematic diversity within the 11 stakeholder-generated clusters reflects a complex implementation landscape for AI diagnostic decision support tools. Notably, several statements sorted during the clustering phase, especially those falling within clusters such as "Trust and Accountability Issues" and "Infrastructure and Integration Barriers", pointed to shared concerns about algorithmic transparency, data integration challenges, and user proficiency requirements. These concerns were independently emphasized by multiple participants across stakeholder groups, suggesting a convergence of perspectives on the technical and organizational conditions necessary for effective AI deployment. Our clustering results demonstrate a conceptual interrelatedness among the individually identified barriers. In other words, through our clusters, the study team and the participants could co-demonstrate that barriers are not isolated but rather interact across institutional and workflow contexts.

*E) Map Integration*

Based on the map interpretation results, we proceeded with map integration, developed pattern match plots, and generated Go-Zone plots. We also determined the 'minimum meaningful' and 'maximum meaningful' number of clusters for each map and identified the most representative statements for each cluster, as follows.

**1) Pattern Match Analysis Between AI Users and AI Developers**

Pattern match graphs were employed to compare cluster-level average ratings between AI users and AI developers across five evaluation dimensions (Desire to Use, Importance, Means to Use, Method to Use, and Reasons to Use). Each graph depicts the relative ranking of clusters for both groups, with the slope of the connecting lines indicating areas of alignment or divergence, as shown in Fig. 5 and elaborated in the following section. i) Desire to Use: Notable divergences emerged for clinical distrust and concerns over AI decision-making (Cluster 10), which AI users ranked highest, but AI developers placed substantially lower, and lack of training and familiarity with AI tools (Cluster 6), which AI developers prioritized more highly. AI users rated infrastructure and integration barriers (Cluster 1) higher than AI developers, whereas AI developers gave greater weight to trust and accountability issues in AI-generated reporting (Cluster 4). ii) Importance: Differences were observed in the prioritization of Insufficient Adaptability to clinical complexity (Cluster 3), ranked among the most important by AI users but placed lower by AI developers. Conversely, AI developers elevated lack of training and familiarity with AI tools (Cluster 6) and trust and accountability issues in AI-generated reporting (Cluster 4) relative to AI users. Both groups assigned mid-tier rankings to bias, reliability, and quality challenges in AI performance (Cluster 11). iii) Means to Use: Regarding practical capability, AI developers rated lack of training and familiarity with AI tools (Cluster 6) and workflow disruption and lack of standardization (Cluster 7) more positively than AI users. Conversely, AI users assigned higher ratings to infrastructure and integration barriers (Cluster 1) and bias, reliability, and quality challenges in AI performance (Cluster 9). Workforce resistance due to job security concerns (Cluster 8) remained consistently low for both groups. iv) Method to Use: A pronounced divergence appeared for clinical distrust and concerns over AI decision-making (Cluster 10), which AI developers ranked much higher than AI users in terms of methodological suitability. AI users rated trust and accountability issues in AI-generated reporting (Cluster 4) lower than AI developers. Both groups ranked bias, reliability, and quality challenges in AI performance (Cluster 9) toward the lower end. v) Reasons to Use: AI developers prioritized trust and accountability issues in AI-generated reporting (Cluster 4) and bias, reliability, and quality challenges in AI performance (Cluster 9) more strongly than AI users, who instead placed greater emphasis on clinical distrust and concerns over AI decision-making (Cluster 10). Workforce resistance due to job security concerns (Cluster 8) received low ratings from both groups, suggesting limited perceived motivation to address this issue compared to others.

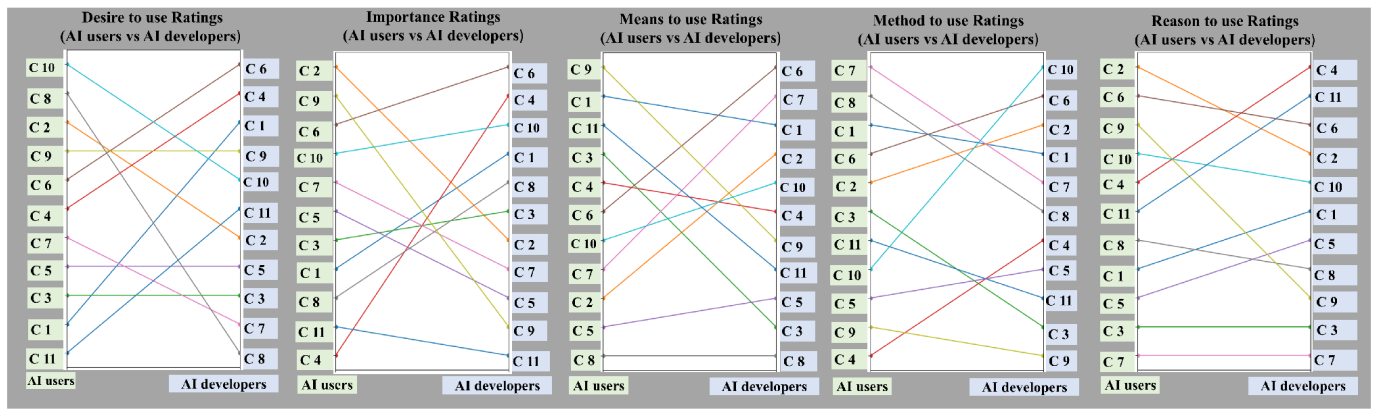

**Fig. 5.** Pattern match graphs comparing cluster (C)-level average ratings between AI users (right axis) and AI developers (left axis) across five evaluation dimensions: Desire to Use, Importance, Means to Use, Method to Use, and Reasons to Use. Each line represents a cluster, with the slope indicating differences in relative ranking between the two groups—steeper slopes denote greater divergence in perceptions, while flatter lines indicate stronger agreement on cluster priorities.

**2) Go-Zone Analysis Results**

In our CM analysis, Go-Zone plots were used to identify priority statements for action by comparing stakeholder ratings across two participant groups—AI users and AI developers. Each point in a plot represents a statement, positioned according to the average ratings provided separately by AI users and AI developers. Following Trochim and McLinden's approach [46], the sub-plot of Fig. 6 is divided into four quadrants using the mean rating for each axis. The upper-right quadrant ("Go-Zone") contains statements rated above average on both dimensions, indicating high-priority, high-readiness actions. The upper-left quadrant shows high-importance but low-feasibility items, representing critical targets for capacity building. The lower-right quadrant contains lower-importance but high-feasibility "quick wins," while the lower-left quadrant includes low-priority items. Go-Zone plots were generated for each thematic cluster of statements, allowing direct comparison between AI users and AI developers. This approach enabled us to pinpoint areas of alignment in priorities as well as points of divergence, providing a clear visual framework for distinguishing actions that can be implemented immediately from those requiring additional skills, resources, or methodological refinement before adoption. We highlight seven primary clusters containing the largest number of Go-Zone (upper-right quadrant) statements, as detailed below, while the remaining four clusters are presented in the supplementary materials (Section 2, Supplemental Fig. S1).

i) Test Acquisition, Integration & Adaptability Gaps (Statements: S1, S3, S4, S6, S7, S9, and S11): Stakeholders identified limited integration of AI tools with imaging and other testing hardware (S11 and S4) and failure to optimize acquisition for complex or atypical cases (S7) as critical barriers. AI systems may also neglect patient comfort (S3) or inadvertently increase scan times (S1), which affects workflow efficiency. Concerns about bias in test quality (S12) further reduce adoption. These issues rank high in "Importance" and "Reasons to Use" but also have relatively high "Desire to Use", suggesting readiness if technical integration improves.

Addressing hardware compatibility, providing technician training (S25), and embedding patient-centered adjustments in protocols are key enablers. ii) Test Ordering, Transparency & Workflow Fit (Statements: S14, S17, S18, S19, S21, S23, S28, S29, and S31): Clinicians emphasized poor integration with electronic health record (EHR) and existing tools (S19, S23, and S29) as a major barrier, which undermines accuracy and efficiency. Lack of transparency in AI-generated orders (S28) and insufficient adaptation to local resource constraints (S17) hinder trust. Training deficits (S21) limit uptake even when "Desire to Use" is high. Legal and ethical concerns (S18 and S31) amplify hesitation. Solutions include improving interoperability, clarifying legal accountability, and enhancing data completeness across facilities. iii) Interpretation Trust and Transparency in AI Analysis (Statements: S44 and S43): Clinicians may distrust AI-driven test interpretation, particularly if results lack transparency (S44) or do not align with clinical intuition (S43). These rank high in "Importance" and "Reasons to Use", with moderate "Desire to Use", suggesting that explainability and alignment with human reasoning are essential for adoption. iv) Test Protocoling, High-Priority Cluster – Patient-Specific Adaptability (Statements: S47, S49, S50, S55, S56, and S66): AI tools often fail to adapt protocols to patient-specific factors (S47 and S50) and may struggle in rare or atypical cases (S56 and S66). Clinicians may resist such AI protocols due to autonomy concerns (S55). These issues fall into the high-importance/high-desire zone but have feasibility constraints in "Means" and "Method" dimensions, indicating the need for adaptive algorithm design and clinician oversight mechanisms. v) Report Communication, Maintaining Clinician-to-Clinician Interaction (Statements: S59, S60, S65, and S68): Over-reliance on AI for communication may reduce essential direct clinician interactions (S59), while poor integration into clinical communication systems (S65 and S60) slows delivery of results. Legal uncertainties (S68) further impede trust. These rank high in "Importance" and "Reasons to Use", with feasible "Means" improvements possible through better integration.

vi) Report Generation, Accuracy and Customization Needs (Statements: S33, S62, and S72): AI-generated reports may lack customization for unique cases (S33), undermining clinician trust. Inaccuracies or incompleteness (S62) and ethical concerns about reliance on AI (S72) add to the barrier profile. These are high in "Importance" and "Reasons to Use", but low in "Means", indicating the need for quality assurance pipelines and customizable templates.

vii) Scheduling, Infrastructure Limitations (Statements: S84, S87, S88, S93, S95, and S96): Healthcare systems often lack the technical infrastructure for AI scheduling (S88 and S95), with tools struggling to adapt to real-time changes (S84 and S87). Financial barriers (S93) and privacy concerns (S96) also rank highly. While strong "Reasons to Use" exist, "Means to Use" remain low, requiring both infrastructure investment and compliance frameworks.

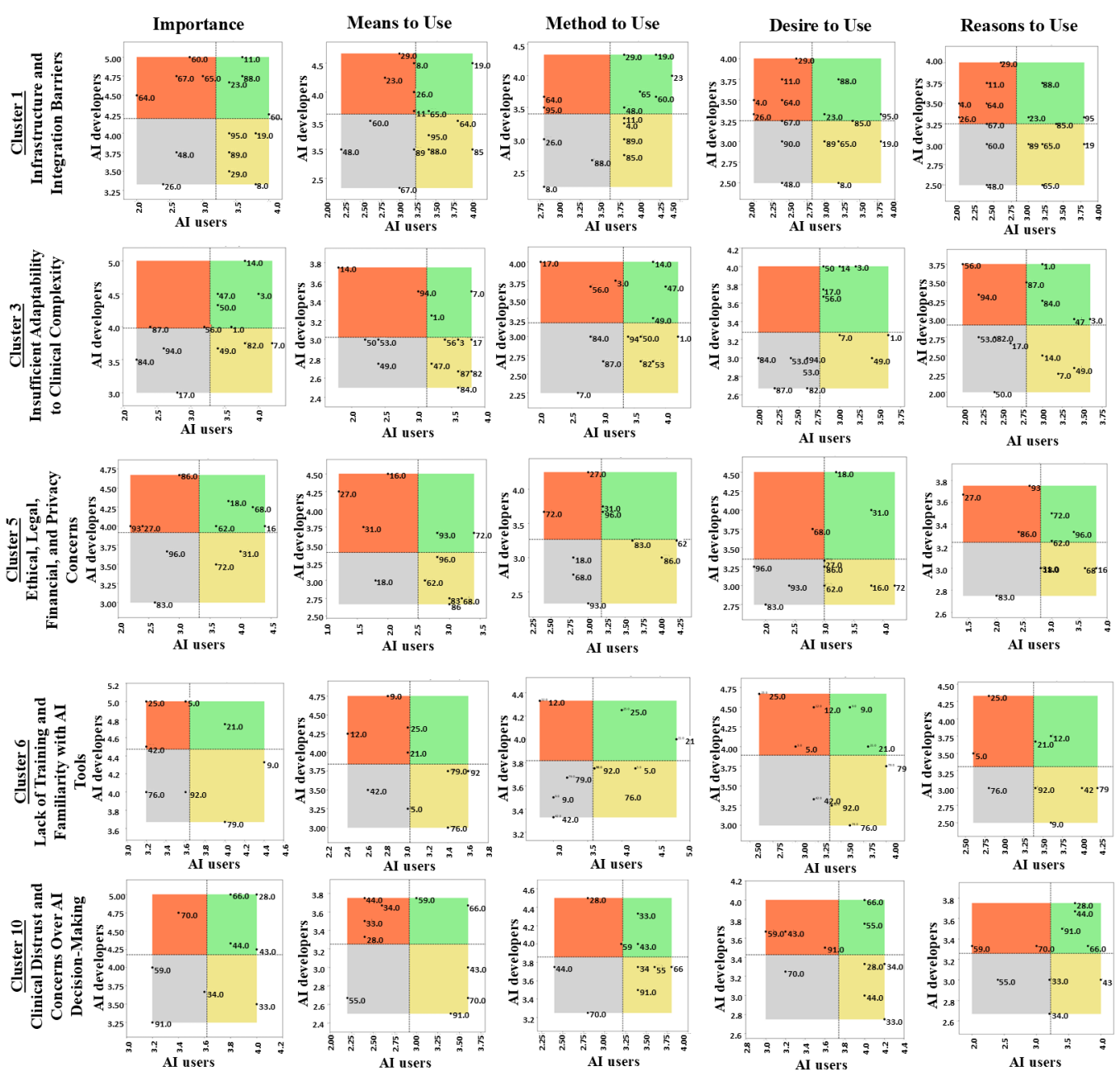

**Fig. 6.** Go-Zone plots for each major cluster across five evaluation dimensions (Importance, Means to Use, Method to Use, Desire to Use, and Reasons to Use), comparing ratings from AI Users (x-axis) and AI Developers (y-axis). Quadrants indicate the relative priority and feasibility of statements within a cluster: upper-right (green) = above-average ratings from both groups ("Go-Zone"); upper-left (orange) = high AI Developer ratings but low AI User ratings; lower-right (yellow) = high AI User ratings but low AI Developer ratings; lower-left (gray) = below-average ratings from both groups.

## IV. DISCUSSION

Participants rated each of the 11 identified clusters across five key dimensions. We discuss selected findings here, focusing on their implications for AI implementation in diagnostic medicine and their potential relevance to emerging healthcare technologies. The integration of AI diagnostic decision support tools into clinical practice offers significant promise for enhancing diagnostic accuracy, reducing delays, and improving workflow efficiency. However, as this study demonstrates through the use of CM within the KTA framework, the path to adoption is fraught with complex, multidimensional barriers. This structured understanding provides a foundation for designing targeted and context-sensitive strategies to support AI integration into clinical practice.

This study provides a foundation for understanding barriers to AI implementation in diagnostic decision support. The stakeholder-sorted statements highlighted challenges related to trust, transparency, data integration, accountability, and usability. Addressing these challenges requires robust technical infrastructure, clinician training, interdisciplinary coordination, and appropriate ethical governance [22]. These barriers may also be relevant to emerging digital twin (DT) applications, which often depend on AI-driven analytics, integrated clinical data, and similar organizational and technical infrastructure. However, further research is needed to determine how these barriers apply specifically to DT implementation. For instance, concerns about accountability in AI-generated reporting highlight the importance of traceability, validation, and clearly defined responsibilities in clinical workflows [49]. Similarly, gaps in training and familiarity, reflected in the "Lack of Training" cluster, indicate broader workforce readiness challenges that may hinder the implementation of AI across clinical departments. Future research should examine how these barriers vary across clinical settings and identify strategies to support safe, transparent, and effective AI adoption [50].

Closely following were "Clinical Distrust in AI Decision-Making" and "Clinical Skepticism and Loss of Human-Centered Decision-Making", reflecting persistent concerns about AI's transparency, explainability, and alignment with human expertise. These barriers are particularly relevant to the integration of AI tools across the diagnostic chain workflow, where AI may support several stages of diagnosis, including image interpretation, risk stratification, differential diagnosis, clinical reporting, and treatment planning. Although emerging evidence suggests that AI can outperform physicians in certain diagnostic tasks, real-world adoption remains uneven [51]. For instance, radiologists have been reported to ignore AI-generated alerts when they conflict with clinical intuition, illustrating that trust remains a cultural and relational challenge rather than merely a technical one [52]. The integration of AI into diagnostic workflows may also have implications for emerging DT applications, as diagnostic data can support patient-specific modeling and prediction. Such applications may benefit from AI-enabled diagnostic information, although their implementation requires further investigation [53-55].

Barriers such as "Infrastructure and Integration" also received significant attention, especially in resource-constrained settings like rural hospitals. Costs associated with infrastructure upgrades pose considerable obstacles for institutions with limited capital and narrow operating margins [56]. This was reflected in stakeholder ratings, where the "Means to Use" and "Method to Use" dimensions scored lower across several clusters, indicating operational and logistical gaps in readiness. Similarly, workflow integration challenges in "Method to Use", emphasize the need for AI tools to fit into existing clinical routines without creating additional burden. AI systems are often not designed with frontline usability in mind, creating friction rather than facilitation in clinical settings [57].

The cluster related to Ethical, Legal, Financial, and Privacy Concerns points to the ongoing need for robust regulatory frameworks that govern liability, data security, consent, and algorithmic transparency [58]. As highlighted by Price et al. [59], the "black-box" nature of many AI models complicates accountability, which in turn hinders clinical confidence and insurer reimbursement models. Bias and equity concerns also surfaced as critical, especially given the potential for AI systems trained on non-representative datasets to underperform in underserved populations such as the elderly, disabled, or homeless individuals [60]. These risks were underscored in a recent report indicating that AI-driven diagnostics may perform worse than random chance when deployed without adequate contextual validation [61].

Implications for Stakeholders: For healthcare providers, the findings emphasize the importance of education and involvement. Clinicians should not only receive training but also be included in AI development and validation processes to ensure alignment with clinical reasoning and usability [62, 63]. Administrators and hospital systems must prioritize investments in digital infrastructure, workflow integration, and staff support. The cost of not addressing these barriers may be higher in terms of failed implementations, clinician burnout, and missed opportunities for innovation [64, 65]. Policymakers play a pivotal role in shaping a regulatory ecosystem that supports innovation while safeguarding patient rights. Clear guidance on liability, data sharing, and reimbursement is essential to reduce institutional hesitancy [66]. AI developers must prioritize adaptability, transparency, and equity. Tools need to be clinically relevant, context-aware, and designed with frontline users in mind. Developing with rather than for clinicians is essential [58].

If addressed effectively, overcoming these barriers could lead to more accurate, timely, and equitable diagnosis, particularly in high-volume or resource-constrained environments. AI integration in diagnostic decision support tools could serve as cognitive extenders, reducing clinician burden and enhancing patient outcomes. However, in line with what has been stated elsewhere [47, 66] and as our stakeholders noted, AI must be positioned as a complement, not a replacement, for human judgment, especially in complex or ambiguous cases.

Limitation: One key limitation of this study is the relatively small number of participants who completed each phase of the CM process. This limits the statistical generalizability of the findings and may have reduced the diversity of perspectives captured. However, participants were purposefully selected because of their direct involvement in AI, diagnostic decision support, and digital transformation-related research, development, and implementation activities. All participants worked within similar project environments, teams, and clinical innovation contexts at BC Cancer, which provided a shared understanding of the practical challenges involved in integrating AI users, clinicians, developers, and implementation teams. This shared context supported an in-depth exploration of barriers grounded in real-world experience. Nevertheless, future research should include larger and more diverse samples across multiple institutions, clinical domains, and health system contexts to validate and extend the current findings.

Future Research Directions: This study provides a foundation for future research on AI implementation in diagnostic decision support. Further studies should examine how barriers related to trust, transparency, data integration, accountability, and usability influence AI adoption across different clinical settings. Although the present study did not directly examine DT implementation, some of the identified barriers may also be relevant to DT-enabled healthcare applications, particularly those relying on AI-based prediction, clinical data integration, and continuous model refinement. Future research should investigate these potential connections and assess the specific implementation requirements of DT technologies in clinical settings.

Future studies should also more explicitly define and examine specific categories of diagnostic AI technologies, as implementation barriers may vary across modalities and tasks. For example, barriers may differ between computer vision applications for medical image classification, detection, or segmentation and large language model (LLM)-based applications for report generation, summarization, or diagnostic decision support. Future work could also focus on specific medical specialties or disciplines, as barriers to AI-supported diagnostic decision-making may vary depending on clinical workflow, disease area, patient population, regulatory requirements, and the maturity of digital and data infrastructure within each specialty. Where feasible, stratified analyses across AI technology categories, clinical settings, and user groups would help distinguish barriers that are specific to particular AI applications from those that reflect broader challenges in healthcare implementation. In addition, future work should examine the wider organizational, technical, and infrastructural conditions required for the successful development and implementation of AI in diagnostic medicine. These include data quality and interoperability, governance arrangements, model validation and monitoring, stakeholder engagement, clinical integration, workforce readiness, and the availability of robust digital health infrastructure.

Future research may explore the use of LLMs and multi-agent systems to support IS tasks, such as synthesizing stakeholder input, identifying implementation barriers, and generating potential strategies. However, these approaches should be viewed as complementary to, rather than replacements for, qualitative methods such as CM. CM offers important advantages because it is participatory, transparent, and grounded in stakeholder perspectives, making it a safer and more context-sensitive method for clinical implementation research. In contrast, LLM- and agent-based approaches may improve efficiency and scalability, but they also face challenges related to bias, reliability, explainability, data privacy, accountability, and acceptance in clinical settings [67]. Therefore, future studies should evaluate how these emerging technologies can be responsibly integrated with established qualitative methods while ensuring safety, trustworthiness, and clinical relevance [68].

The various insights from our clusters can guide practical strategies for planning and improving implementation. Each cluster points to specific actions, such as creating targeted training programs, developing workflow integration tools, or setting up ethical review processes, that could also support AI adoption in diagnostic medicine. For example, participatory design workshops can help address workflow disruptions, while stakeholder discussions on privacy and consent can create clear, trusted processes [70]. These approaches are scalable and can be adapted to different health systems. The concept maps from our study give developers and implementation teams a clear, evidence-based guide for setting priorities and making decisions. By following these maps, teams can focus on the most important barriers first and work in ways that reflect the needs and input of stakeholders.

Overall, our findings contribute to a broader methodological and translational foundation for implementing AI diagnostic decision support tools in clinical practice. IS offers an approach that aligns technological advances with human-centered design and organizational feasibility. The stakeholder-informed knowledge generated through our CM and KTA exercise may help guide the development of context-sensitive implementation strategies. Longitudinal studies can evaluate the impact of specific interventions over time. Comparative studies across different healthcare settings can help identify context-specific vs. generalizable barriers. Given the rapid pace of AI developments, periodic reassessment of implementation barriers through updated CM exercises is recommended. Social media discussions and practitioner-generated content should also be considered as real-time, qualitative data sources that reflect frontline perspectives and emerging concerns.

## V. Conclusion

This study presents a comprehensive, stakeholder-driven analysis of barriers to AI diagnostic decision support tools adoption in diagnostic medicine. By applying CM within the KTA framework, we identified and prioritized 11 critical barriers across technical, cultural, ethical, and systemic domains. These insights offer a structured foundation for designing targeted, context-sensitive strategies that bridge the gap between AI innovation and clinical application. While our main focus was on AI diagnostic decision support workflows, the overlap between the identified barrier clusters and the infrastructure and ethical issues linked to DTs points to wider implications for bringing these technologies into practice. As DT technologies continue to evolve—especially those using AI for personalized radiopharmaceutical therapies and predictive imaging—the stakeholder-informed clusters presented here provide a practical framework that IS can apply in different settings for medical testing and imaging. By illuminating shared challenges and intervention strategies, our findings support strategic integration of both AI diagnostic decision support tools and DT systems in future clinical research and real-world deployment. This positions our work as a resource for developers, policymakers, and implementers striving for context-aware, equitable, and scalable adoption of digital innovations, particularly in medical diagnostics.

## Acknowledgment

We gratefully acknowledge the support of the Technological Virtual Collaboration (TECVICO Corp.), the Virtual Collaboration Group (VirCollab, www.vircollab.com), Vancouver, Canada, and Tehran University of Medical Sciences. The authors also thank the experts and stakeholders who participated in the surveys and interviews, including N. Sanati, S. Gharibi, I. Alberts, C. Gowdy, M. Asadi, S. Falahati, S. Dehghanfard, S. Jabarzadeh, S. Kurkowska, G. Mousavi, M. Maghsudi, S. Amiri, S. S. Mehrnia, H. Garivani, H. Ashayeri, M. M. Mehrabi Nejad, A. M. Ahmadzadeh, S. J. Qandilu, and S. M. Piri. Moreover, all authors declare that they have no known conflicts of interest in terms of competing financial interests or personal relationships that could have an influence or are relevant to the work reported in this paper.

CONFLICT OF INTEREST

Mohammad Salmanpour is affiliated with TECVICO CORP. The remaining co-authors declare no relevant conflicts of interest.